\documentclass[11pt]{article}
\usepackage{amsfonts,amssymb,amsmath,amsthm,enumerate,empheq}
\def\address#1{\par
    {\centering{\footnotesize\it#1\par}}\par\vspace*{11pt}}
\def\keywords#1{\par
    \vspace*{8pt}
    {\footnotesize{\leftskip18pt\rightskip\leftskip
    \noindent{\it Keywords}\/:\ #1\par}}\vskip-12pt}
\newtheorem{theorem}{Theorem}[section]

\let\refcite\cite
\let\cites\cite

\def\e{\mathrm{e}}
\def\i{\mathrm{i}}
\def\d#1{\mathrm{d}{#1}}
\DeclareMathOperator{\esssup}{ess\,sup}
\usepackage{bbm}

\begin{document}

\title{Three useful bounds in quantum mechanics ---
 easily obtained by Wiener integration}

\author{Hajo LESCHKE$^*$ and Rainer RUDER}
\date{}

\maketitle

\address{Institut f\"{u}r Theoretische Physik, Universit\"{a}t Erlangen-N\"{u}rnberg,\\
91058 Erlangen, Germany\\
$^*$E-mail: hajo.leschke@physik.uni-erlangen.de\\
URL: www.theorie1.physik.uni-erlangen.de}

\vspace{-9cm}\hspace{-3cm}
\begin{minipage}{\textwidth}
\scriptsize {\it To be published in shortened and different form in the proceedings:} 

Path Integrals --- New Trends and Perspectives,

edited by W.~Janke and A.~Pelster (World Scientific, Singapore, 2008) 

See also: arXiv:0810.2495 [quant-ph]
\end{minipage}
\vspace{+9cm}

\begin{abstract}
In a reasonably self-contained and explicit presentation we
illustrate the efficiency of the Feynman--Kac formula for the rigorous
derivation of three inequalities of interest in non-relativistic
quantum mechanics.
\end{abstract}


\begin{center}
{\it ``As it was, as soon as I heard Feynman describe his path integral approach to quantum mechanics during a lecture at Cornell everything became clear at once; and there remained only matters of rigor.''} \cite{k79}
\end{center}
\begin{center}
{\it ``Hardly a month passes without someone discovering yet another application \emph{[of the Feynman--Kac formula]}. Remarkable and, of course, highly gratifying.''} \cite{k85}
\end{center}
\begin{center}
 Mark~Kac~(1914\,--\,1984)
\end{center} 

\section{Feynman--Kac formula}\label{secfk}
In order to model the phenomenon of Brownian motion Norbert Wiener~\cite{w23} introduced as early as 1923 a certain probability distribution
$\mu$ concentrated on the set $W^d$  of continuous paths
$ w: [0,\infty[ \rightarrow \mathbb{R}^d, t\mapsto w(t)$
from the positive half-line $[0,\infty[\subset\mathbb{R}^1$ into $d$-dimensional
Euclidean space $\mathbb{R}^d$, $d\in\{1,2,3,\dots\}$, which start at the origin, $w(0)=0$. This
distribution is a mathematically well-defined positive measure in the sense of
general measure theory \cite{r87} and therefore induces a corresponding concept
of integration over paths, which we denote by $\int_{W^d}\!\mu(\d{w})
(\boldsymbol{\cdot})$.
The (standard) \emph{Wiener measure} $\mu$ is uniquely determined by requiring that it
is \emph{Gaussian} with normalization
\begin{equation}
\label{zerom}
\boxed{\int_{W^d}\!\mu(\d{w}) =1}
\end{equation}
and first, respectively, second moments given by
\begin{equation}\label{moment}
\boxed{
\int_{W^d}\!\mu(\d{w})\, w_j(t)=0,\qquad\int_{W^d}\!\mu(\d{w})\,
w_j(t)w_k(s)=\delta_{jk}\min\{t,s\}
}
\end{equation}
for all $j,k\in\{1,\dots,d\}$ and all $t,s\in [0,\infty[$.
Here $w_j(t)$ denotes the $j$-th component of the path $w$
evaluated at (time) parameter $t\geq 0$. The simple Wiener integrals 
(\ref{zerom}) and (\ref{moment}) imply that the components of $w$ are, 
in probabilistic language, centered, independent and
identically distributed. 

To Wiener integrals apply, in contrast to Feynman path integrals, all
the rules and computational tools provided by general measure and
integration theory, most notably Lebesgue's dominated-convergence
theorem~\cite{r87}. As a consequence, Wiener integration often serves, via the
Feynman--Kac formula, as an efficient technique for obtaining
results in quantum mechanics with complete rigour. An impressive
compilation of such results was given by Barry Simon already in
1979. Since then not much has changed which is reflected by the 
fact that the second edition of his book \cite{s05} differs from the
first one only by an addition of bibliographic notes on some of the more recent
 developments. Still Wiener integration should be considered
neither as a secret weapon nor as a panacea for obtaining rigorous results in
quantum mechanics. In any case, the Feynman--Kac formula is more than just
a poetic rewriting of a Lie--Trotter formula. Ironically, Richard
Feynman himself took advantage of that as early as 1955 in his
celebrated paper on the polaron \cite{f55}, in particular, by using the Jensen
inequality.

Now, what is the Feynman--Kac formula? Let us consider a spinless
charged particle with configuration space $\mathbb{R}^d$ subjected to
a scalar potential $v: \mathbb{R}^d\rightarrow \mathbb{R}^1, q\mapsto
v(q), q=(q_1,\dots,q_d)$ and a vector potential
$a:\mathbb{R}^d\rightarrow \mathbb{R}^d, q\mapsto a(q),
a=(a_1,\dots,a_d)$. The latter generates a magnetic field (tensor)
defined by $b_{jk}:=\partial a_k/\partial q_j-\partial a_j/\partial
q_k$. The corresponding (non-relativistic) quantum system is
informally given by the Hamiltonian
\begin{equation}
\boxed{
H(a,v):=\big(P-a(Q)\big)^2/2 + v(Q)
}
\end{equation} where
$Q=(Q_1,\dots,Q_d)$ and $P=(P_1,\dots,P_d)$ denote the  $d$-component
operators of position and canonical momentum, respectively. They obey
the canonical commutation relations $Q_j P_k - P_k Q_j =\text{i}\hbar
\delta_{jk}\mathbbm{1}$. Here $\i=\sqrt{-1}$ is the imaginary unit and $\hbar >0$ is Planck's constant (divided by $2\pi$). Moreover, we have chosen physical units where both the mass and the charge of the particle are equal to $1$. Under
rather weak assumptions \cite{blm04} on $a$ and $v$, $H(a,v)$
can be defined as a self-adjoint operator on the Hilbert space
$\text{L}^2(\mathbb{R}^d)$ of all (equi\-valence classes of) Lebesgue
square-integrable complex-valued functions on $\mathbb{R}^d$. Furthermore 
its ``Boltzmann--Gibbs operator'' $\e^{-\beta H(a,v)}$ even posesses for
each $\beta\in]0,\infty[$ an integral kernel $\langle q |\e^{-\beta H(a,v)} |
q^\prime \rangle$ (in other words, position representation or
Euclidian propagator) which is jointly continuous in $q,q^\prime\in\mathbb{R}^d$.

We are now prepared to state the \emph{Feynman--Kac
formula}. Apart from mathematical subtleties its content is most concisely
 expressed by the following representation free version:
\begin{empheq}[innerbox=\fbox,
right=
\parbox{0.5cm}{
\vspace{1.32cm}
\hspace{0.03cm}.
}
]
{multline}\label{fk}
\e^{-\beta H(a,v)}=\int_{W^d}\!\mu(\d{w})\,\e^{-\i w(\beta\hbar^2)\boldsymbol{\cdot} P/\hbar}
\exp\Big\{-\!\!\int_0^\beta\!\! \d{\tau}\, v(w(\tau\hbar^2)\mathbbm{1}+Q)\Big\}
\\\times\exp\Big\{\i\hbar\int_0^\beta\!\!\!\d{\tau} \dot w(\tau\hbar^2)\boldsymbol{\cdot} a\big(w(\tau\hbar^2)\mathbbm{1}+Q\big)\Big\}
\end{empheq}
Here the dot ``\,$\boldsymbol{\cdot}$\,'' between two $d$-component quantities refers to the Euclidian scalar product of $\mathbb{R}^d$
and the integral containing the vector potential is a suggestive notation for a \emph{stochastic line
integral} in the sense of R.\,L.~Stratonovich and D.\,L.~Fisk (corresponding to a mid-point
discretization). In (\ref{fk}) the Wiener integration serves to
disentangle the non-commuting operators $P$ and $Q$ in $\e^{-\beta
H(a,v)}$.
For related remarks see Ref.~\refcite{blw96} and references therein. We recall from (\ref{moment}) that $\mu$ neither depends on $\beta$ nor on~$\hbar$. Also $P/\hbar$ is independent of~$\hbar$.

By going
informally to the position representation of (\ref{fk}) one gets the rigorously proven formula 
\begin{multline}\label{kernel}
\!\langle q |\e^{-\beta H(a,v)}|q^\prime \rangle=\int_{W^d}\!\!\mu(\d{w})\delta\big(w(\beta\hbar^2)+q^\prime-q\big)
\exp\Big\{-\int_0^\beta\!\!\d{\tau} v\big(w(\tau\hbar^2)+q^\prime\big)\Big\}\\
\times\exp\Big\{\i\hbar\int_0^\beta\! \d{\tau}\, \dot w(\tau\hbar^2)\boldsymbol{\cdot}
a\big(w(\tau\hbar^2)+q^\prime\big)\Big\}.
\end{multline}
It even holds for a class of potentials $v$ for which $H(a,v)$ is not
bounded from below \cite{blm04} and remains true (by the self-adjointness of $H(a,v)$), if on its right side  $q$ and $q^\prime$ are exchanged and simultaneously $\i$ is changed to $-\i$. The Dirac delta in (\ref{kernel})
indicates that all paths $w$ to be integrated over arrive
in $q-q^\prime \in\mathbb{R}^d$ at ``time'' $\beta\hbar^2 > 0$. In fact, they may be considered to end
there, because $\mu$ is Markovian and the Wiener integrand in
(\ref{kernel}) does not depend on $w(\tau\hbar^2)$ for $\tau >\beta$. More
precisely, the path integration may  be performed with
respect to the \emph{Brownian bridge} \cites{s05,blm04}. In the Appendix below we shall present what we think is an illuminating derivation of the ``bridge version'' of (\ref{kernel}), although in the following we shall not make (explicit) use of that version.

In the next three sections we are going to illustrate the usefulness of
(\ref{kernel}) by deriving three inequalities of interest in quantum
mechanics.


\section{Diamagnetic inequality}\label{secdia}

\begin{theorem}
\begin{equation}\label{dia}
\left| \langle q |\e^{-\beta H(a,v)}|q^\prime \rangle\right|
\leq\langle q |\e^{-\beta H(0,v)}|q^\prime \rangle 
\end{equation}
holds for all $\beta > 0$ and all $q, q^\prime \in\mathbb{R}^d$.
\end{theorem}
\begin{proof} Inequality (\ref{dia}) is an immediate consequence of (\ref{kernel}) by
taking the absolute value, applying the ``triangle inequality''
\begin{equation}
\left|\int_{W^d}\!\mu(\d{w})\,\left(\boldsymbol{\cdot}\right)\right|\leq 
\int_{W^d}\!\mu(\d{w})\,\left|\left(\boldsymbol{\cdot}\right)\right|
\end{equation}
and using the elementary
identity $\left|\e^{x+\i y}\right|=\e^{x}$ for $x, y\in\mathbb{R}^1$. 
\end{proof}
Remarks:
\begin{enumerate}[(i)]

\item This elegant proof is due to Edward Nelson, see Ref.~\refcite{s06}
(also for other historical aspects of (\ref{dia}) and related inequalities). 

\item If $\int_{\mathbb{R}^d}\! \d{q}\,\e^{-\beta v(q)}<\infty$, the \emph{free
energy} $-\beta^{-1}\ln\int_{\mathbb{R}^d}\!\d{q}\,\langle q
|\e^{-\beta H(0,v)}|q \rangle$ at inverse temperature $\beta >
0$ exists and (\ref{dia}) then implies that it cannot be lowered
by  turning on a magnetic field. Under weaker assumptions on
$v$, for example for the hydrogen atom (that is, for $d=3$ and
$v(q)=-\gamma/|q|$ with $\gamma > 0$) (\ref{dia}) still implies in the limit $\beta\rightarrow \infty$
the same sort of stability for the \emph{ground-state energy}.
Altogether this explains the name \emph{diamagnetic inequality}.

\item  There are also diamagnetic inequalities in case the particle is
restricted to a region in $\mathbb{R}^d$ of finite volume with Dirichlet, Neumann
or other boundary conditions \cites{hlmw01a,hs04}.
Moreover, the proof of the diamagnetic inequality easily extends to
the case of many (interacting) particles, provided there is no spin
and no Fermi statistics involved.
\end{enumerate}
An interesting question is what can be said if $a\neq 0$ is changed
(pointwise) to another vector potential $a^\prime\neq
0$. For a partial answer see Sec.~\ref{secmono} below.


\section{Quasi-classical upper bound on the integrated density of
states in the case of a random scalar potential}

In the single-particle theory of electronic properties of disordered
or amorphous solids the scalar potential $v$ in $H(a,v)$ is considered to be a realization of a \emph{random} field on
$\mathbb{R}^d$ which is distributed according to some probability measure
$\nu$ on some set $V$ of potentials $v$.  We denote by $\int_V\!\nu(\d{v})\,(\boldsymbol{\cdot})$ the
corresponding (functional) integration or averaging. One example 
is a \emph{Gaussian} $\nu$ with vanishing first moments and second
moments given by
$\int_V\!\nu(\d{v})\,v(q)v(q^\prime)=C(q-q^\prime)$
for all $q,q^\prime\in\mathbb{R}^d$ with some (even) covariance function
$C:\mathbb{R}^d\rightarrow\mathbb{R}^1$. The fact that the second
moments only depend on the difference $q-q^\prime$ reflects the
assumed ``homogeneity on average''. We also assume that $C$ is
continuous, $C(q)$ tends to zero as $|q|\rightarrow
\infty$ and the single-site variance obeys $0<C(0)<\infty$. The
$\mathbb{R}^d$-homogeneity together with the decay of the correlations
of the fluctuations at different sites with increasing
distance implies the \emph{$\mathbb{R}^d$-ergodicity} of the 
(Gaussian) random potential.

A quantity of basic interest in the above-mentioned theory is the \emph{integrated density of states}. It
may be defined \cites{pf92,lmw03,blm04} as the non-decreasing function  
$N:\mathbb{R}^1\rightarrow\mathbb{R}^1,E\mapsto N(E,a,q)$ where 
\begin{equation}\label{ids}
N(E,a,q):=\int_V\!\nu(\d{v})\,\langle q |\Theta(E-H(a,v))|q \rangle.
\end{equation}
Here $\Theta$ denotes Heaviside's unit-step function and the (non-random) vector
potential $a$ as well as the position $q\in\mathbb{R}^d$ are considered as parameters.
If the random potential (characterized by $\nu$) and the magnetic field
(generated by $a$) are both homogeneous, then $N(E,a,q)$ actually does
 not depend on $q$. 
Of course, in the physically most relevant cases the random potential
should be even ergodic, so that $N(E,a,0)$ coincides for $\nu$-almost all
 realizations $v$ with the number of eigenvalues per volume of a
finite-volume restriction of $H(a,v)$ below the energy
$E\in\mathbb{R}^1$ in the infinite-volume limit. 
Nevertheless, the following estimate holds also for random potentials and
magnetic fields which are not homogeneous. 

\begin{theorem}\label{th2}
If the probability measure $\nu$ of the random potential has the
property\footnote{Given a function $f:\mathbb{R}^d\rightarrow\mathbb{R}^1$, then
$\esssup_{r\in\mathbb{R}^d}|f(r)|$ denotes the smallest 
$M\in[0,\infty]$ such that $|f(r)|\leq M$ holds for Lebesgue-almost
all $r\in\mathbb{R}^d$.}
that  $L_\beta:=\esssup_{r\in\mathbb{R}^d}\int_V\!\nu(\d{v})\,\e^{-\beta
v(r)}<\infty$ for all $\beta > 0$, then 
\begin{equation}\label{pas}
N(E,a,q)\leq (2\pi\beta\hbar^2)^{-d/2} L_\beta\, \e^{\beta E}
\end{equation}
holds for all energies $E\in\mathbb{R}^1$ and all $\beta > 0$.
\end{theorem}

\begin{proof}
\begin{align}
& N(E,a,q)\,\e^{-\beta E}\leq \int_V\!\nu(\d{v})\,\langle q |\e^{-\beta
H(a,v)}| q \rangle 
\leq \int_V\!\nu(\d{v})\,\langle q |\e^{-\beta H(0,v)}| q \rangle \label{p2:ele} \\
& = \int_{W^d}\!\mu(\d{w})\,\delta\big(w(\beta\hbar^2)\big)\int_V\!\nu(\d{v})\,\exp\Big\{-\int_0^\beta\!
\d{\tau}\, v(w(\tau\hbar^2)+q)\Big\}  \label{p2:fk} \\
&\leq \int_0^\beta\!\frac{\d{\tau}}{\beta}\int_{W^d}\!\mu(\d{w})\,\delta\big(w(\beta\hbar^2)\big)\int_V\!\nu(\d{v})\,\e^{-\beta
v(w(\tau\hbar^2)+q)}\\
& \leq \left(2\pi\beta\hbar^2\right)^{-d/2} L_\beta.
\label{p2:jen}
\end{align}
Here (\ref{p2:ele}) is due to the elementary inequality
$\Theta(E-H(a,v))\leq \e^{\beta(E-H(a,v))}$, referring to the spectral theorem,
 and (\ref{dia}).
Eq.~(\ref{kernel}) then gives (\ref{p2:fk}).
The next inequality is Jensen's with respect to the uniform average 
$\beta^{-1}\int_0^\beta\!\d{\tau}\,(\boldsymbol{\cdot})$. The claim
now follows from the definition of $L_\beta$, Eq.~(\ref{kernel}) with
$(a,v)=(0,0)$ and $\int_0^\beta\!\d{\tau}=\beta$.
The various interchanges of integrations can be justified by the Fubini-Tonelli
theorem \cite{r87}.
\end{proof}
Remarks:
\begin{enumerate}[(i)]

\item Theorem \ref{th2} is a slight extension of a result which goes
back to Pastur, see  Thm.~9.1  in Ref.~\refcite{pf92}.
The right side of (\ref{pas}) is \emph{quasi-classical} in
the sense that it does not depend on $a$ and does not take into
account, due to the  Jensen inequality (\ref{p2:jen}) in its proof,
the non-commutativity of the kinetic and potential energy.

\item While the estimate (\ref{pas}) holds for rather general random
potentials,
 the various inequalities in its proof are responsible for
its roughness, even when optimized with respect to $\beta >
0$. Nevertheless, it shows that $N(E)$  decreases to $0$ at least exponentially
fast as $E\rightarrow-\infty$. For a homogeneous Gaussian random
potential (and a constant magnetic field) the optimized estimate even
reflects the exact Gaussian decay \cites{pf92,lmw03} $\ln N(E)\sim -E^2/2C(0)$  as
$E\rightarrow-\infty$. For a non-Gaussian random potential, like a
repulsive Poissonian one, the leading low-energy decay of $N$, the
\emph{Lifshitz tail}, typically is of true quantum nature \cites{pf92,lmw03,hw04} and
can therefore not be reflected by the right side of
(\ref{pas}). Although the (universal) leading high-energy growth
$N(E)\sim (E/2\pi\hbar^2)^{d/2}/\Gamma(1+d/2)$ as $E\rightarrow\infty$, see
Refs.~\refcite{pf92,lmw03} and references therein, is quasi-classical,
the optimized right side of (\ref{pas}) overestimates it slightly by a constant factor 
(due to the elementary inequality in (\ref{p2:ele})).

\item For related  quasi- and pseudo-classical bounds (with a non-random $v$) we refer to Sec.~9 in Simon's book \cite{s05} and to Ref.~\refcite{lw89}.

\end{enumerate}


\section{ A simple diamagnetic monotonicity}\label{secmono}
For a partial answer to the question raised at the end of
Sec.~\ref{secdia} we only consider the planar case $d=2$ with
$v=0$ and a perpendicular magnetic field not depending 
on the second co-ordinate $q_2$. We assume that $b:=b_{12}$ is a
continuously differentiable function of $q_1\in\mathbb{R}^1$. One possible vector
potential generating $b$, not depending on $q_2$ either, is given by $a^{(b)}(q):=\big(0,\int_0^{q_1}\!\d{r}\,
b(r)\big)$.
In the following theorem $H^{(b)}$ denotes any Hamiltonian on
$\mathrm{L}^2(\mathbb{R}^2)$ which is gauge equivalent to 
$H(a^{(b)},0)=P_1^2/2+\big(P_2-a^{(b)}_2(Q_1)\big)^2/2$ for the given $b$. The assertion
(\ref{thm3}) is therefore gauge invariant.
Nevertheless, in the proof we will use $H(a^{(b)},0)$ and see
that one can dispense with the absolute value  on the right side of (\ref{thm3})
in this particular gauge.
\begin{theorem}\label{mono}
If $b$ and $B$ are two magnetic fields as just described  and satisfy either
$\left|b(r)\right|\leq B(r)$ or $\left|b(r)\right|\leq -B(r)$ for all $r\in\mathbb{R}^1$, then
\begin{equation}\label{thm3}
\left|\langle q |\e^{-\beta H^{(B)}}|q^\prime
\rangle\right|\leq\left|\langle (q_1,0) |\e^{-\beta
H^{(b)}}|(q_1^\prime,0)\rangle\right|\e^{-(q_2-q_2^\prime)^2/(2\beta\hbar^2)}
\end{equation}
holds for all $\beta > 0$, all $q=(q_1,q_2) \in\mathbb{R}^2$ and all $q^\prime=(q^\prime_1,q^\prime_2) \in\mathbb{R}^2$.
\end{theorem}
\begin{proof}
By (\ref{kernel}) the left side of (\ref{thm3}) is invariant under a global sign change of
$B$. Therefore it suffices to consider the case $B(r)\geq 0$.
For notational transparency we put $\hbar=1$ and write $a$ and $A$ for $a_2^{(b)}$ and
$a_2^{(B)}$, respectively. For a given pair $(\beta,w)\in ]0,\infty[ \times W^1$  we introduce the notations
\begin{align}
m_\beta(a,w)&:=\beta^{-1}\int_0^\beta\!\!\d{\tau}\,a(w(\tau)),\\
s^2_\beta(a,w)&:=\beta^{-1}\int_0^\beta\!\!\d{\tau}\,\big(a(w(\tau))\big)^2
-\big(m_\beta(a,w)\big)^2
\end{align}
for the mean and variance of $a\big(w(\tau)\big)$  with respect to the uniform
average $\beta^{-1}\int_0^\beta\!\d{\tau}\,(\boldsymbol{\cdot})$, and similarly with $A$ instead of $a$. 
Next we observe the following two ``doubling identities''
\begin{align}
&2\beta^2\big[s^2_\beta(A,w)-s^2_\beta(a,w)\big]\\
&=\int_0^\beta\!\!\d{\tau}\int_0^\beta\!\!\d{\sigma}\Big\{\big[A(w(\tau))-A(w(\sigma))\big]^2-\big[a(w(\tau))-a(w(\sigma))\big]^2\Big\}\\
&=\int_0^\beta\!\!\d{\tau}\int_0^\beta\!\!\d{\sigma}\big[a_+(w(\tau))-a_+(w(\sigma))\big]\big[a_-(w(\tau))-a_-(w(\sigma))\big].
\end{align}
The last integrand is non-negative, because the two functions $r\mapsto a_\pm(r):=A(r)\pm a(r)=\int_0^r\d{r^\prime}\big(B(r^\prime) \pm b(r^\prime)\big)$, $r\in\mathbb{R}^1$,
are both non-decreasing since $B(r^\prime)\geq|b(r^\prime)|\geq\mp
b(r^\prime)$ by assumption. The same arguments apply when the path
$w$ is replaced by the  rigidly shifted one $w+q_1^\prime\in W^1+\mathbb{R}^1$ defined by $(w+q_1^\prime)(\tau):=w(\tau) + q_1^\prime$. To
summarize, we have shown so far that
\begin{equation}\label{vardiff}
s^2_\beta(a,w+q_1^\prime)\leq s^2_\beta(A,w+q_1^\prime).
\end{equation}

Since $H^{(B)}$, in the particular gauge chosen,  commutes with $P_2$, we have
\begin{equation}\label{decomp1}
H^{(B)}=\int_{\mathbb{R}^1}\d{k} H^{(B)}(k)\otimes|k\rangle\langle k|,
\end{equation}
using an informal
notation for a direct-integral decomposition. Here the one-parameter family of effective Hamiltonians
\begin{equation}\label{decomp2}
H^{(B)}(k):=P_1^2/2+\big(k\mathbbm{1}-A(Q_1)\big)^2/2,\quad k\in\mathbb{R}^1,
\end{equation}
acts on the Hilbert space $\textrm{L}^2(\mathbb{R}^1)$ of (wave) functions of the first co-ordinate.
By (\ref{decomp1}) and (\ref{decomp2}) we get
\begin{align}
&\langle q| \e^{-\beta H^{(B)}}|q^\prime\rangle =(2\pi)^{-1}\int_{\mathbb{R}^1}\d{k}\langle q_1|\e^{-\beta
H^{(B)}(k)}|q_1^\prime\rangle \,\e^{\i k(q_2-q_2^\prime)}\\
\begin{split}
&=(2\pi\beta)^{-1/2}\e^{-(q_2-q_2^\prime)^2/(2\beta)}\int_{W^1}\mu(\d{w})\delta(w(\beta)+q_1^\prime-q_1)\\
&\quad \times \exp\Big\{-\beta s^2_\beta(A,w+q_1^\prime)/2\Big\}
\exp\Big\{\i(q_2-q_2^\prime)m_\beta(A,w+q_1^\prime)\Big\}
\label{gl1}.
\end{split}
\end{align}
Here we have used (\ref{kernel}) with $d=1$, $a=0$
and $v=(k-A)^2/2$ and then performed the (Gaussian)
integration with respect to $k$. By applying the ``triangle
inequality'' to (\ref{gl1}) and then using (\ref{vardiff}) we finally obtain

\begin{align}
\begin{split}
&\e^{(q_2-q_2^\prime)^2/(2\beta)}\left|\langle q| \e^{-\beta
H^{(B)}}|q^\prime\rangle\right|\\ 
&\leq (2\pi\beta)^{-1/2}\int_{W^1}\!\mu(\d{w})\delta(w(\beta)+q_1^\prime-q_1)
\exp\Big\{-\beta s^2_\beta(a,w+q_1^\prime)/2\Big\}
\end{split}\\
&=\langle (q_1,0)| \e^{-\beta H^{(b)}}|(q_1^\prime,0)\rangle=\left|\langle (q_1,0)| \e^{-\beta H^{(b)}}|(q_1^\prime,0)\rangle\right|.
\end{align}
The last two equalities follow again from (\ref{gl1}) with $a$ instead of $A$.
\end{proof}
Remarks:
\begin{enumerate}[(i)]
\item To our knowledge, Theorem \ref{mono} first appeared in
Ref.~\refcite{lrw02}. It complements some of the results obtained by Loss, Thaller and
Erd\H{o}s \cites{lt97,e97}. For a survey of results of this
genre see Sec.~9 in Ref.~\refcite{rm05}.

\item For a given sign-definite $B$ the right side of (\ref{thm3}) can
be made explicit by choosing for $b$ the globally constant field
$B_0:=\inf_{r\in\mathbb{R}^1}|B(r)|$, so that
\begin{equation*}
\left|\langle q| \e^{-\beta
H^{(B)}}|q^\prime\rangle\right|
\leq\frac{B_0/4\pi\hbar}{\sinh(\beta\hbar
B_0/2)}\exp\!\Big\{-
\frac{(q_1-q_1^\prime)^2B_0/4\hbar}{\tanh(\beta\hbar B_0/2)}-\frac{(q_2-q_2^\prime)^2}{2\beta\hbar^2}\Big\}.
\end{equation*}
If $B_0\neq 0$, the Gaussian decay on the right side is faster along
the 1- than along the 2-direction. Such an anisotropy has  been found
also for
the almost-sure transport properties in the case that $B$ is a
 (Gaussian) random field with non-zero mean \cite{lww06}.

\end{enumerate}

\begin{appendix}
\section*{Appendix}
For convenience of the reader we are going to derive the bridge version of the Feynman--Kac formula (\ref{kernel}). We start out from a fixed triple $(T,q,q^\prime)\in\,]0,\infty[\,\times\mathbb{R}^d\times\mathbb{R}^d$ and associate to each continuous path $w:[0,\infty[\rightarrow\mathbb{R}^d$ with $w(0)=0$ another path $\widehat{w}: [0,T]\rightarrow\mathbb{R}^d$ defined by
\begin{equation}\label{bridgepath}
\widehat{w}(t):=w(t)+q^\prime-\frac{t}{T}\big(w(T)+q^\prime-q\big),\qquad t\in[0,T].
\end{equation}
Obviously, $\widehat{w}$ is a \emph{bridge path} in the sense that $\widehat{w}\in \Omega^d_{T,q,q^\prime}$ where 
\begin{equation}
\Omega^d_{T,q,q^\prime}:=\left\{\omega:[0,T]\rightarrow\mathbb{R}^d \,\big|\, \omega \textrm{ is continuous, }\omega(0)=q^\prime, \omega(T)=q\right\}
\end{equation}
is the set of all continuous paths connecting position $q^\prime$ to position $q$ in the time period $T$. In fact, $\Omega^d_{T,q,q^\prime}$ is the image of $W^d$ under the mapping $w\mapsto \widehat{w}$ given by (\ref{bridgepath}).
\newpage\noindent
Writing $\langle (\cdot)\rangle:=\int_{W^d}\!\mu(\d{w}) (\cdot)$ for the expectation or averaging induced by the Wiener measure, Eqs.~(\ref{zerom}) and (\ref{moment}) yield
\begin{align}\label{bmom1}
&\langle{\widehat{w}_j(t)}\rangle=q_j^\prime-\frac{t}{T}(q_j^\prime-q_j),\\
&\langle{\widehat{w}_j(t)\widehat{w}_k(s)}\rangle-\langle{\widehat{w}_j(t)}\rangle\langle{\widehat{w}_k(s)}\rangle=\delta_{jk}\Big(\min\{t,s\}-\frac{ts}{T}\Big)\label{bmom2}
\end{align}
for the first and second moments of the bridge paths (\ref{bridgepath}), in other words, for their mean and covariance.

Hence, by the (affine) linearity and the surjectivity of the mapping $w\mapsto\widehat{w}$ the induced image of the Wiener measure $\mu$ on $W^d$ is a \emph{Gaussian} probability measure $\rho_{T,q,q^\prime}$ on $\Omega_{T,q,q^\prime}^d$  with mean and covariance given by the right sides of (\ref{bmom1}) and (\ref{bmom2}).
Other useful consequences of (\ref{moment}) and (\ref{bridgepath}) are the two equalities
\begin{equation}
\langle w_k(T)\widehat{w}_j(t)\rangle=0=\langle w_k(T)\rangle\langle\widehat{w}_j(t)\rangle.
\end{equation}
They hold for all $j,k\in\{1,\dots,d\}$ and all $t\in[0,T]$ and imply, by the Gaussian nature of the Wiener measure $\mu$,  that the random point $w(T)\in\mathbb{R}^d$ is independent of the family of random points $\{\widehat{w}(t)\}_{t\in[0,T]}\subset\mathbb{R}^d$.

Now, how can all this be applied to the right side of (\ref{kernel})? First we get from (\ref{bridgepath}) that
\begin{equation}
\delta(w(T)+q^\prime-q)F_T(w+q^\prime)=\delta(w(T)+q^\prime-q)F_T(\widehat{w})
\end{equation}
for any complex-valued function(al) $F_T$ on $W^d+\mathbb{R}^d$, not depending on the points $w(t)$ of $w$ for $t> T$. The two mentioned independencies then give
\begin{align}
\langle\delta(w(T)+q^\prime-q)F_T(w+q^\prime)\rangle
&=\langle\delta(w(T)+q^\prime-q)\rangle\langle F_T(\widehat{w})\rangle\\
&=(2\pi T)^{-d/2}\e^{-(q-q^\prime)^2/2T}\langle F_T(\widehat{w})\rangle
\end{align}
and the \emph{bridge version} of the Feynman--Kac formula (\ref{kernel}) eventually reads
\begin{multline}
\langle q |\e^{-\beta H(a,v)}|q^\prime \rangle
=(2\pi\beta\hbar^2)^{-d/2}\,\e^{-(q-q^\prime)^2/2\beta\hbar^2}\\
\times\int_{\Omega_{\beta\hbar^2,q,q^\prime}^d}\!\!\!\!\!\!\rho_{\beta\hbar^2,q,q^\prime}(\d{\omega})
\,\exp\Big\{-\int_0^\beta\!\d{\tau} v\big(\omega(\tau\hbar^2)\big)\Big\}\\
\times
\exp\Big\{\i\hbar\int_0^\beta\! \d{\tau}\, \dot{\omega}(\tau\hbar^2)\boldsymbol{\cdot}a\big(\omega(\tau\hbar^2)\big)\Big\}.
\end{multline}
Remaining matters of rigour can be supplied.

We note that a ``one-parameter decomposition'' similar to (\ref{bridgepath}) for a Gaussian random potential with non-negative covariance function has turned out to be useful in proving local Lipschitz continuity of the corresponding integrated density of states (\ref{ids}), see the proof of Cor.~4.3 in Ref.~\cite{hlmw01a}.
\end{appendix}

\bibliographystyle{hieeetr}
\bibliography{bounds}

\end{document}